\documentclass[5p]{elsarticle}

\pdfoutput=1

\usepackage[T1]{fontenc}
\usepackage[latin9]{inputenc}
\usepackage[english]{babel}
\usepackage{amsmath}
\usepackage{amssymb}
\usepackage{wasysym}
\usepackage{graphicx}
\usepackage{esint}
\usepackage[unicode=true,
 bookmarks=false,
 breaklinks=false,pdfborder={0 0 1},backref=false,colorlinks=true,citecolor=blue,urlcolor=blue,linkcolor=blue]
 {hyperref}
\usepackage{todonotes}

\makeatletter

\usepackage{hyperref}
\usepackage{bbm}
\usepackage{cleveref}
\usepackage{bbold}

\usepackage{feynmp}
\newcommand{\Mp}{M_\text{P}}
\newcommand{\mn}{{\mu\nu}}
\newcommand{\ab}{{\alpha\beta}}
\newcommand{\dl}{\xrightarrow{\text{DL}}}

\newcommand{\dd}[2][]{\, \text{d}^{#1}{#2} \,}
\newcommand{\dv}[3][]{\frac{\text{d}^{#1}{#2}}{\text{d}{#3}^{#1}}}

\title{Beyond dRGT -- a new massive gravity theory?\tnoteref{t1}}

\tnotetext[t1]{Declarations of interest: none}

\begin{document}

\author[itp]{Manuel Wittner\corref{cor1}}
\ead{wittner@thphys.uni-heidelberg.de}
\selectlanguage{english}%

\author[itp]{Frank K\"onnig}
\ead{koennig@thphys.uni-heidelberg.de}

\author[sbu]{Nima Khosravi}
\ead{n-khosravi@sbu.ac.ir}

\author[itp]{Luca Amendola}
\ead{amendola@thphys.uni-heidelberg.de}

\cortext[cor1]{Corresponding author}

\address[itp]{Institut f\"{u}r Theoretische Physik, Ruprecht-Karls-Universit\"{a}t Heidelberg,
Philosophenweg 16, 69120 Heidelberg, Germany}

\address[sbu]{Department of Physics, Shahid Beheshti University, G.C., Evin, Tehran 19839, Iran}

\date{\today}

\begin{abstract}
It is commonly believed that the dRGT theory is the unique way to describe a massive spin-2 field without ghosts. While dRGT is arguably the most elegant massive gravity theory, it seems that it may not be the unique ghost-free one if one relaxes the assumption of locality. In this work, we derive a new massive gravity theory by using a disformal transformation of the metric tensor in the dRGT action. Its decoupling limit lives inside the class of beyond-Horndeski Lagrangians as long as the transformation of the metric remains purely disformal. This proves the absence of ghosts in this decoupling limit and hints at their absence in the whole theory. One caveat, however, is the possible nonlocal structure of this new theory. Furthermore, we consider a more general case, in which we allow the conformal factor in the disformal transformation to be different from unity, and discuss the absence of ghosts in this decoupling limit.
\end{abstract}

\begin{keyword}
modified gravity, massive gravity, dRGT, massive graviton, beyond Horndeski
\end{keyword}

\maketitle

\section{Introduction}

The search for a viable theory with certain properties is greatly simplified by the knowledge about a unique class of theories. A prominent example is General Relativity (GR) which was shown to be the unique theory that is solely constructed from a rank-2 tensor with the following properties: it is symmetric, divergence-free, only second order in the derivatives of $g$, linear in these derivatives and it naturally depends on a pseudo-Riemannian metric $g$ \cite{Vermeil:1917,Cartan:1922}. Since GR describes a massless spin-2 field, it is natural to ask whether the graviton is theoretically allowed to be massive.
\par
At linear level, the search for theories describing a massive spin-2 field is conceptionally rather simple. The action can only consist of a linear combination of a finite number of terms. The absence of ghosts can then be used to constrain the prefactors. This finally results in the Fierz-Pauli theory, the unique linear theory of a Massive Gravity (MG) \cite{Fierz:1939ix}.
\par
Generalizing the Fierz-Pauli action towards a non-linear theory has turned out to be rather cumbersome. Boulware and Deser had thought to find a ghost in the whole class of theories of MG \cite{Boulware:1973my}. However, a loophole in their arguments has been used to reveal a stable non-linear theory of MG \cite{deRham:2010ik,deRham:2010kj,Hassan:2011vm,Hassan:2011hr,Hassan:2011tf,Hassan:2011ea,Creminelli:2005qk,Hassan:2012qv}, often dubbed as dRGT theory.
\par
Even though dRGT is stable, it cannot provide dynamical cosmological solutions \cite{DAmico:2011eto} and is therefore not considered to be viable. Does this imply that a MG without any modifications, like, e.g., breaking the cosmological principle \cite{DAmico:2011eto}, assuming Lorentz violation \cite{Rubakov:2004eb,Konnig:2016idp}, or adding a new dynamical tensor field \cite{Hassan:2011ea,Hassan:2011zd}, cannot exist? This would certainly be the case if dRGT was unique. Even though this is commonly assumed to be the case, a proof is still missing. In fact, in this work we will present a counter example by constructing a new theory of MG. This \textit{beyond dRGT massive gravity} turns out to be ghost-free in the Decoupling Limit (DL) which strongly hints at the stability of the theory at full non-linear level. Furthermore, the finding of another theory of MG questions the uniqueness of dRGT and possibly opens the door to a whole class of theories of a massive spin-2 field. However, we emphasize that this theory might have nonlocalities in the non-linear regime.

\section{Formalism}
\subsection{Construction of Beyond dRGT}

To construct a new ghost-free theory of MG, we start with dRGT MG in the perturbative approach \cite{deRham:2010ik}:
\begin{equation}
	\mathcal{L} = \Mp^2 \sqrt{-g} \left[ R - \frac{m^2}{4} \sum_i U_i(g,H)\right], \label{eq_lagrangian_untransformed}
\end{equation}
where the functions $U_i$ are of the $i$-th order on the covariantized metric perturbation $H$. In the $\Lambda_3$-decoupling limit (DL), this Lagrangian becomes \cite{deRham:2010ik}
\begin{equation}
	\mathcal{L}^{\Lambda_3}_\text{dRGT} = - \frac{1}{2} h^{\mu \nu} \mathcal{E}^{\alpha \beta}_{\mu \nu} h_{\alpha \beta} + h^{\mu \nu} \bar{X}^{(N)}_{\mu \nu} (\pi), \label{eq_lagrangian_untransformed_DL}
\end{equation}
where $h$ is the non-covariantized metric perturbation and $\bar{X}^{(N)}_{\mu \nu}$ is a superposition of first and higher powers of second derivatives of the helicity-0 mode $\pi$. There are strong indications that all powers $> 3$ of these second derivatives vanish \cite{deRham:2010ik}.

In Ref. \cite{deRham:2010ik}, it has been shown that the Lagrangian \eqref{eq_lagrangian_untransformed_DL} can be reformulated in terms of the Galileon Lagrangians, which live inside the class of Horndeski Lagrangians. The latter ones can be mapped to the Beyond-Horndeski (BH) Lagrangians through a Disformal Transformation (DT) \cite{Gleyzes:2014qga}. The idea of this work is to revert the order of these steps, beginning with a disformally transformed dRGT theory: if the DL of this theory lives inside BH, we have found a proof for the absence of ghosts in it.

We start with a DT of the metric tensor in the Lagrangian \eqref{eq_lagrangian_untransformed}, which is given by \cite{Bekenstein:1992pj, Bettoni:2013diz, Zumalacarregui:2013pma, Koivisto:2012za}
\begin{equation}
	g_{\mu \nu} \rightarrow \tilde{g}_{\mu \nu} \equiv C(X,\phi) g_{\mu \nu} + D (X,\phi) \phi_{, \mu} \phi_{, \nu}, \label{eq_definition_disformal_transformation}
\end{equation}
where $C$ and $D$ are arbitrary functions and \\ $X \equiv - \frac{1}{2} g^{\mu \nu} \phi_{, \mu} \phi_{, \nu}$ is the kinetic term of a new scalar degree of freedom (DoF) $\phi$.
The resulting Lagrangian,
\begin{equation}
	\tilde{\mathcal{L}} = \Mp^2 \sqrt{-\tilde{g}} \left[ \tilde{R} - \frac{m^2}{4} \sum_i U_i(\tilde{g},\tilde{H}) \right], \label{eq_action_transformed}
\end{equation}
describes a new theory of MG, where the matter sector remains untransformed, i.e., the matter fields couple minimally to $g_\mn$. Here we replaced $\tilde{U}_i$ by $U_i$ because it keeps the same form as in the untransformed case with only the arguments replaced by transformed quantities. In analogy to dRGT MG \cite{deRham:2010ik}, these transformed quantities are given by
\begin{align}
	\tilde{g}_{\mu \nu} &\equiv \eta_{\mu \nu} + \tilde{h}_{\mu \nu}/\Mp \equiv \tilde{H}_{\mu \nu} + \eta_{ab} \pi^a_{;\mu} \pi^b_{;\nu} \label{eq_metric_perturbation_transformed} \\
	\Rightarrow \tilde{H}_{\mu \nu} &= \frac{\tilde{h}_{\mu \nu}}{\Mp} + \frac{2}{\Mp m^2} \tilde{\Pi}_{\mu \nu} - \frac{1}{\Mp^2 m^4} \tilde{\Pi}^2_{\mu \nu},
\end{align}
with
\begin{equation}
	\frac{\tilde{h}_{\mu \nu}}{\Mp} = -\eta_{\mu \nu} + C (\eta_{\mu \nu} + \frac{h_{\mu \nu}}{\Mp}) + D\phi_{,\mu} \phi_{,\nu}. \label{eq_h_transformed}
\end{equation}
Here, $\tilde{\Pi}_\mn \equiv \tilde{\nabla}_\mu \tilde{\nabla}_\nu \pi$ is the second covariant derivative of the scalar mode with respect to the transformed metric. It is important to mention that the DT \eqref{eq_definition_disformal_transformation} introduces a new scalar DoF $\phi$. In order to preserve the number of DoFs, we will from now on identify this new scalar field with the helicity-0 mode of MG, i.e. $\phi \equiv \pi$.\footnote{Note that in a general case, the helicity-0 mode $\pi$ and hence also the new scalar DoF $\phi$ depend on the background, which however does not imply that the number of DoFs is changed.}

The transformation of the metric determinant and Ricci scalar leads to \cite{Zumalacarregui:2013pma,Zumalacarregui:2012us}
\begin{align}
	\tilde{g} &= C^3(C-2DX)g, \\
	\tilde{R} &= \tilde{g}^{\mu \nu} \left( R^\alpha_{\;\;\mu \alpha \nu} - 2 \mathcal{K}^\alpha_{\;\;\gamma [\alpha} \mathcal{K}^\gamma_{\;\;\mu] \nu} \right),
\end{align}
where $\mathcal{K}^\alpha_{\;\; \mu \nu} = \tilde{\Gamma}^\alpha_{\mu \nu} - \Gamma^\alpha_{\mu \nu}$ denotes the difference of the transformed and untransformed connections. The transformed inverse metric is given by
\begin{align}
	\tilde{g}^{\mu \nu} = A \left( \eta^{\mu \nu} - \frac{h^{\mu \nu}}{\Mp} \right) - E \phi^{,\mu} \phi^{,\nu}, \label{eq_inverse_metric_transformed}
\end{align}
with $E(X,\phi) \equiv - \frac{B(\tilde{X},\phi)}{A(\tilde{X},\phi) - 2 B(\tilde{X},\phi) \tilde{X}}$ and $A$, $B$, and $\tilde{X}$ defined in Ref. \cite{Zumalacarregui:2013pma}.

\subsection{Constraints on $C$ and $D$} \label{sec_constraints_on_C_and_D}

Surprisingly, an explicit calculation of the interaction terms will not yield a proper DL. For the second order interaction we obtain
\begin{align}
	\Mp^2 m^2 U_2(\tilde{g}, \tilde{H}) &= \Mp^2 m^2 \tilde{g}^\mn \tilde{g}^\ab \left( \tilde{H}_{\mu \alpha} \tilde{H}_{\nu \beta} \right. - \left. \tilde{H}_\mn \tilde{H}_\ab \right) \nonumber \\
	&= \tilde{g}^{\mu \nu} \tilde{g}^{\alpha \beta} \left[ m^2(\tilde{h}_{\mu \alpha} \tilde{h}_{\nu \beta} - \tilde{h}_{\mu \nu} \tilde{h}_{\alpha \beta}) \right. \nonumber \\
	&\qquad + \frac{2}{M_{Pl}m^2}(-\tilde{h}_{\mu \alpha} \tilde{\Pi}^2_{\nu \beta} + \tilde{h}_{\mu \nu} \tilde{\Pi}^2_{\alpha \beta}) \nonumber \\
	&\qquad + \left.\ 4(\tilde{h}_{\mu \alpha} \tilde{\Pi}_{\nu \beta} - \tilde{h}_{\mu \nu} \tilde{\Pi}_{\alpha \beta}) + \vphantom{\frac{2}{\Mp m^2}} \mathcal{O}(\tilde{\Pi}^3) \right].
\end{align}
Since $\tilde{g}^\mn$ contains terms $\mathcal{O}(1) + \mathcal{O}(1/\Mp)$ while $\tilde{h}_\mn$ contains terms $\mathcal{O}(1) + \mathcal{O}(\Mp)$, we obtain a large variety of mass scales if we expand the above expression.
\par We can see that, no matter which mass scale $\Mp^\kappa m^\lambda$ we keep fixed in the DL, as soon as $\Mp \dl \infty$, we obtain divergent terms. This implies that we would need two cutoff scales for our theory to be valid. In order to avoid these complications from the beginning and to render our theory viable at low energy scales, we demand that $\tilde{h}_\mn$ stays constant in the DL. In order to achieve this, we rescale the functions $C$ and $D$ in Eq. \eqref{eq_h_transformed} with powers of $\Mp$ in such a way, that $\tilde{h}_\mn$ does not change with $\Mp$.\footnote{Notice that we could also rescale these functions with powers of $m$, however, in this analysis we restrict ourselves to the assumption that $C$ and $D$ are independent of $m$.} The demand of a fixed $\tilde{h}_\mn$ translates into the condition, that the derivative of Eq. \eqref{eq_h_transformed}, considered as a function of the Planck mass $\tilde{h}_\mn (C(\Mp), D(\Mp),\Mp)$, vanishes:
\begin{align}
	\dv{\tilde{h}_{\mu \nu}}{\Mp} &= (\Mp \eta_{\mu \nu} + h_{\mu \nu}) \dv{C}{\Mp} + (C-1)\eta_{\mu \nu} \nonumber \\
	&\qquad + \Mp \phi_{,\mu} \phi_{,\nu} \dv{D}{\Mp} + D \phi_{,\mu} \phi_{,\nu} \overset{!}{=} 0. \label{eq_constraints_on_C_and_D}
\end{align}
In the following, we will consider three special cases:
\begin{itemize}
	\item[(a)] \textbf{Trivial Case}: Neither $C$ nor $D$ depend on the Planck mass. In this case, Eq. \eqref{eq_constraints_on_C_and_D} leads to a contradiction except if one sets $C=1$ and $D \phi_{,\mu} \phi_{,\nu} = 0$ because $\eta_\mn$ and $\phi_{,\mu} \phi_{,\nu}$ cannot be proportional to each other. These choices for $C$ and $D$ correspond to not performing a transformation at all.
	\item[(b)] \textbf{Purely Disformal Case (PDC)}: Only $D$ depends on $\Mp$ while $C$ does not. We will see that this assumption forces $C$ to equal 1, which corresponds to a `purely disformal transformation'. This case will render a valid and ghost-free theory.
	\item[(c)] \textbf{Generalized Disformal Case (GDC)}: Both, $C$ and $D$, depend on $\Mp$. We will have a look at a specific example of this case, namely that of a `rather general disformal transformation', although it is unclear so far whether the resulting theory is stable.
\end{itemize}

\subsection{Purely Disformal Case}

\subsubsection{Calculating the Metric}

With the assumptions in this case, Eq. (\ref{eq_constraints_on_C_and_D}) becomes
\begin{equation}
	(C-1)\eta_{\mu \nu} + \Mp \phi_{,\mu} \phi_{,\nu} \dv{D}{\Mp} + D \phi_{,\mu} \phi_{,\nu} = 0. \label{eq_constraints_on_C_and_D_2}
\end{equation}
With the same argumentation as in the trivial case, we run into a contradiction if $C \neq 1$, so that Eq. \eqref{eq_constraints_on_C_and_D_2} translates into the differential equation
\begin{equation}
	\dv{D(X,\phi,\Mp)}{\Mp} = - \frac{D(X,\phi,\Mp)}{\Mp},
\end{equation}
with the simple solution $D(X,\phi,\Mp) = \Mp^{-1}D_{0}(X,\phi)$, where the index denotes that $D_0$ is a new arbitrary function independent of $\Mp$.\footnote{Actually, $D_0$ appears in the solution as an integration constant. Strictly speaking, this means that it must not depend on $X$, because $X$ is a contraction with the metric, which depends on $\Mp$. However, this dependence vanishes in the limit $\Mp \dl \infty$, so we will keep the $X$ dependence in $D_0$ to cover a more general case. The same argumentation holds for the free functions in the Eqs. \eqref{eq_constraints_on_C_3}, \eqref{eq_constraints_on_D_3}, \eqref{eq_constraints_on_A_3} and \eqref{eq_constraints_on_E_3}.} Together with the constraint $C(\phi,X) = 1$ this guarantees that $\tilde{h}_{\mu \nu}$ remains constant in the DL. Plugging in these two constraints into Eq. \eqref{eq_h_transformed} and using the definition of $\tilde{h}_\mn$ \eqref{eq_metric_perturbation_transformed}, we find
\begin{align}
	{ \vphantom{g} }^\text{p} \tilde{g}_{\mu \nu} = \eta_{\mu \nu} + \frac{1}{\Mp} (h_{\mu \nu} + D_0 (X,\phi) \phi_{,\mu} \phi_{,\nu}), \label{eq_g_transformed_2}
\end{align}
where the prescript denotes that these are quantities related to the PDC.

The corresponding expression for the inverse transformed metric is obtained by performing an analogue analysis. Therefore we rescale the free functions $A$ and $E$ in Eq. \eqref{eq_inverse_metric_transformed} with powers of $\Mp$ where for the purely disformal case we demand that $E$ depends on the Planck mass while $A$ does not. This leads to the constraints $A=1$ and $E = E_0 / \Mp$ so that the metric is given by
\begin{align}
	{\vphantom{g}}^\text{p} \tilde{g}^{\mu \nu} = \eta^{\mu \nu} - \frac{1}{\Mp} (h^{\mu \nu} + E_0(X,\phi) \phi^{,\mu} \phi^{,\nu}). \label{eq_g_inverse_transformed_2}
\end{align}

\subsubsection{The Decoupling Limit}

The Lagrangian \eqref{eq_action_transformed} consists of two parts: a transformed Einstein-Hilbert term and a transformed interaction term. In order to calculate the DL of the first one, we use that the metric goes to Minkowski and hence the connections of the metric vanish. The DL of the interaction term can be obtained by an analogous calculation as in the dRGT case. At each order some terms can be combined into total derivatives with respect to the transformed metric so that the DL is given by
\begin{equation}
	\frac{\Mp^2 m^2}{4} \sqrt{-{ \vphantom{g} }^\text{p} \tilde{g}} \sum_i U_i ({ \vphantom{g} }^\text{p} \tilde{g}, { \vphantom{H} }^\text{p} \tilde{H}) \dl { \vphantom{h} }^\text{p} \tilde{h}^\mn \tilde{\bar{X}}^{(N)}_{\mu \nu} (\pi),
\end{equation}
where the tilde above $\bar{X}^{(N)}_{\mu \nu} (\pi)$ indicates that it is a function of the second covariant derivatives with respect to the transformed metric ${\vphantom{g}}^\text{p} \tilde{g}_\mn$. Since in the DL the metric becomes Minkowski, these covariant derivatives reduce to partial derivatives. Plugging in the expression for ${\vphantom{h}}^\text{p} \tilde{h}^\mn$ we arrive at the full $\Lambda_3$-DL
\begin{equation}
	{\vphantom{\mathcal{L}}}^\text{p} \tilde{\mathcal{L}}^{\Lambda_3} = \mathcal{L}^{\Lambda_3}_\text{dRGT} + {\vphantom{\mathcal{L}}}^\text{p} \tilde{\mathcal{L}}^{\Lambda_3}_\text{EH} + {\vphantom{\mathcal{L}}}^\text{p} \tilde{\mathcal{L}}^{\Lambda_3}_\text{MG},
\end{equation}
with
\begin{align}
	{\vphantom{\mathcal{L}}}^\text{p} \tilde{\mathcal{L}}^{\Lambda_3}_\text{EH} &= \frac{1}{2} D_0^2(X,\phi) \left\{ \langle \Pi^2 \rangle - \langle \Pi \rangle [\Pi] \vphantom{\frac{1}{2}} \right. \nonumber \\
	&\qquad + \left. \frac{1}{2} (\partial \phi)^2 ([\Pi]^2 - [\Pi^2]) \right\}, \\
	{\vphantom{\mathcal{L}}}^\text{p} \tilde{\mathcal{L}}^{\Lambda_3}_\text{MG} &= D_0 (X,\phi) \phi^{,\mu} \phi^{,\nu} \bar{X}^{(N)}_{\mu \nu} (\pi). \label{eq_action_transformed_DL_2_detailed}
\end{align}
Here we used the same conventions as in Ref. \cite{Zumalacarregui:2013pma}, where $[ \cdot ]$ denotes the trace of the matrix inside the brackets and $\langle \cdot \rangle$ its contraction with the first derivatives $\phi^{,\mu} \phi^{,\nu}$ of the scalar field $\phi \equiv \pi$. Furthermore, $\Pi_\mn$ is the second derivative of this scalar field.

\subsubsection{Searching for Ghosts}

The BH Lagrangians represent the most general 4-dimensional scalar-tensor theory in a curved spacetime that is free of Ostrogradsky instabilities. By showing that the DL of our new theory lives inside this class, we have immediately shown the absence of ghosts. For that purpose, we need to show that we can reproduce each term of the DL by the BH Lagrangians at the same time. Eq. \eqref{eq_action_transformed_DL_2_detailed} consists of the ordinary, ghost-free dRGT term $\mathcal{L}^{\Lambda_3}_\text{dRGT}$ plus a new term ${\vphantom{\mathcal{L}}}^\text{p} \tilde{\mathcal{L}}^{\Lambda_3}_\text{EH}$ from the transformed Einstein-Hilbert action and a new term ${\vphantom{\mathcal{L}}}^\text{p} \tilde{\mathcal{L}}^{\Lambda_3}_\text{MG}$ from the transformed interaction terms.

$\mathcal{L}^{\Lambda_3}_\text{dRGT}$ can be formulated in terms of Galileons \cite{deRham:2010ik}, which are basically the Horndeski Lagrangians in flat space and hence lives inside BH.

Tuning the free functions $G_i (X,\phi)$ and $F_i (X,\phi)$ of the BH Lagrangians as defined in Ref. \cite{Gleyzes:2014dya}, we can reproduce the first of the new terms ${\vphantom{\mathcal{L}}}^\text{p} \tilde{\mathcal{L}}^{\Lambda_3}_\text{EH}$.

The second of the new terms consists of three terms:
\begin{equation}
	{\vphantom{\mathcal{L}}}^\text{p} \tilde{\mathcal{L}}^{\Lambda_3}_\text{MG} = D_0 \phi^{,\mu} \phi^{,\nu} \left( \bar{X}^{(1)}_{\mu \nu} + \frac{1}{\Lambda_3^3} \bar{X}^{(2)}_{\mu \nu} \right. + \left. \frac{1}{\Lambda_3^6} \bar{X}^{(3)}_{\mu \nu} \right).
\end{equation}
The first of these terms can be reproduced by the Horndeski functions $G_2$ and $G_3$ through the usage of a procedure developed in Ref. \cite{Zumalacarregui:2013pma}. The latter two of these three terms can be shown to live inside BH by tuning the functions $F_4$ and $F_5$. In detail, all of the new terms compared to the dRGT case can be reproduced by the BH Lagrangians by choosing
\begin{align}
	G_2(X,\phi) &= - 2X f_{,\phi}, \label{eq_G2_second_case} \\
	G_3(X,\phi) &= f - 2XD_0, \label{eq_G3_second_case} \\
	F_4(\phi,X) &= \frac{1}{8} D_0^2(X,\phi) - \frac{6c_3-1}{4\Lambda_3^3} D_0(X,\phi), \label{eq_F4_second_case} \\
	F_5 (X,\phi) &= \frac{c_3 + 8d_5}{\Lambda_3^6} D_0(X,\phi), \label{eq_F5_second_case}
\end{align}
where $f(X,\phi) = -\int D_0(X,\phi) dX + s(\phi)$. This proves that the DL of the purely disformal case is free of ghosts.

\subsubsection{Fierz-Pauli limit of the PDC}

In the limit where $\pi_{,\mu} = \phi_{,\mu} = 0$, according to Eq. \eqref{eq_g_transformed_2}, the transformed metric equals the untransformed one:
\begin{equation}
	{ \vphantom{g} }^\text{p} \tilde{g}_{\mu \nu} = g_\mn.
\end{equation}
Hence, we obtain the same linear limit as dRGT, i.e. the Fierz-Pauli limit.

\subsubsection{The Purely Disformal Case in Terms of Galileons}

The new terms in the DL can be rewritten in terms of Galileons:
\begin{align}
	{\vphantom{\mathcal{L}}}^\text{p} \tilde{\mathcal{L}}^{\Lambda_3}_\text{EH} &= \frac{1}{4} D_0^2 \mathcal{L}^\text{Gal}_4, \\
	{\vphantom{\mathcal{L}}}^\text{p} \tilde{\mathcal{L}}^{\Lambda_3}_\text{MG} &= D_0 \left( \mathcal{L}^\text{Gal}_3 - \frac{6c_3-1}{2 \Lambda_3^3} \mathcal{L}^\text{Gal}_4 - \frac{c_3 + 8d_5}{\Lambda_3^6} \mathcal{L}^\text{Gal}_5 \right).
\end{align}
Here we use the same form of the Galileons as in Ref. \cite{Gleyzes:2014qga}, where they are described as an alternative in flat space, or respectively the so called ``form 3'' from Ref. \cite{Deffayet:2013lga}. Unlike the Galileons, these terms are not invariant under a Galilean shift symmetry anymore because of the prefactors $D_0^2$ and $D_0$, respectively. Just like in the DL of dRGT \cite{deRham:2010ik}, by setting $c_3=-8d_5$ the quintic order of the Galileons and the new terms vanishes as well as the mixing term $h^\mn X^{(3)}_\mn$. Thus, the DL of our theory can be written as
\begin{align}
	{\vphantom{S}}^\text{p} \tilde{S}^{\Lambda_3} &= \int \dd[4]{x} \left[ - \frac{1}{2} \hat{h}_{\mu \nu} \hat{\mathcal{E}}^{\mu \nu \alpha \beta} \hat{h}_{\alpha \beta} + \mathcal{L}^\text{Gal} \right. \\
	 &\qquad + \left. D_0 \left\{ \mathcal{L}^\text{Gal}_3 + \left( \frac{1}{4} D_0 - \frac{6c_3-1}{2 \Lambda_3^3} \right) \mathcal{L}^\text{Gal}_4 \right\} \right], \nonumber
\end{align}
where $\mathcal{L}^\text{Gal}$ is a linear combination of the Galileon Lagrangians up to fourth order.

\subsubsection{Nonlocal structure}

The identification $\phi \equiv \pi$ relates the new scalar degree of freedom from the disformal transformation to the helicity-0 mode of the graviton. The latter one is defined on the Minkowski background, which raises the question about the corresponding identification on a different background or in a background-independent formulation. Retaining the identification with the helicity-0 mode on the Minkowski background, we need to find an expression for $\pi$ on a general background. Given the fact that $\tilde{H}_\mn$ contains second derivatives of $\pi$ in the form of $\tilde{\Pi}_\mn$, translating $\pi$ to arbitrary backgrounds in general requires to invert these second derivatives to obtain $\pi$ in terms of the metric components. This inversion induces integrals and hence nonlocal operators into the action. A local description as in dRGT might exist but this is a priori not obvious.

On the Minkowski background, the decoupling limit of the PDC is obviously local. The full theory however is given by the Lagrangian Eq. \eqref{eq_action_transformed}, which contains an infinite power series of second order derivative operators acting on $\pi$. Thus, a nonlocal structure might possibly arise even on the Minkowski background.

\subsection{Generalized Disformal Case}

\subsubsection{Calculating the Metric}

Now we allow both functions $C$ and $D$ to depend on the Planck mass so that the whole Eq. \eqref{eq_constraints_on_C_and_D} holds. One rather generic, yet, not the most general solution can be obtained by setting the prefactors of $\eta_{\mu \nu}$ to zero, i.e.:
\begin{equation}
	\Mp \dv{C(X,\phi,\Mp)}{\Mp} + C(X,\phi,\Mp) - 1 = 0, \label{eq_constraints_on_C_3}
\end{equation}
which can be solved for $C(X,\phi,\Mp) = \Mp^{-1} C_0 (X,\phi) + 1$. Together with Eq. \eqref{eq_constraints_on_C_and_D}, we arrive at
\begin{equation}
	\frac{C_0}{\Mp^2} h_{\mu \nu} = \left( \Mp \dv{D(\Mp)}{\Mp} + D(\Mp) \right) \phi_{,\mu} \phi_{,\nu},
\end{equation}
which yields the following constraint:
\begin{equation}
	\phi_{,\mu} \phi_{,\nu} D(\Mp) = \phi_{,\mu} \phi_{,\nu} \frac{D_0}{\Mp} - h_{\mu \nu} \frac{C_0}{\Mp^2}. \label{eq_constraints_on_D_3}
\end{equation}
With these two constraints of Eqs. (\ref{eq_constraints_on_C_3}) and (\ref{eq_constraints_on_D_3}), which are more general than those from the purely disformal case (we obtain that case if we set $C_0 (X,\phi) = 0$), the transformed metric becomes:
\begin{align}
	{\vphantom{g}}^\text{g} \tilde{g}_{\mu \nu} = \eta_{\mu \nu} + \frac{1}{\Mp} (C_0 \eta_{\mu \nu} + h_{\mu \nu} + D_0 \phi_{,\mu} \phi_{,\nu}), \label{eq_g_transformed_third_case}
\end{align}
where the prescript denotes that these quantities are related to the GDC. An analogous calculation can be performed for the inverse metric, yielding the constraints
\begin{equation}
	A(X,\phi,\Mp) = \frac{A_0(X,\phi)}{\Mp} + 1 \label{eq_constraints_on_A_3}
\end{equation}
and
\begin{align}
	\phi^{,\mu} \phi^{,\nu} E(\Mp) &= \phi^{,\mu} \phi^{,\nu} \frac{E_0}{\Mp} - h^{\mu \nu} \frac{A_0}{\Mp^2}, \label{eq_constraints_on_E_3}
\end{align}
which leads to the following form of the inverse transformed metric:
\begin{align}
	{\vphantom{g}}^\text{g} \tilde{g}^{\mu \nu} = \eta^{\mu \nu} - \frac{1}{\Mp} (-A_0 \eta^{\mu \nu} + h^{\mu \nu} + E_0 \phi^{,\mu} \phi^{,\nu}). \label{eq_g_inverse_transformed_third_case}
\end{align}


\subsubsection{The Decoupling Limit}

Plugging in Eqs. (\ref{eq_g_transformed_third_case}) and (\ref{eq_g_inverse_transformed_third_case}) into Eq. \eqref{eq_action_transformed} and calculating the $\Lambda_3$-DL, we obtain
\begin{equation}
	{\vphantom{\mathcal{L}}}^\text{g} \tilde{\mathcal{L}}^{\Lambda_3} = {\vphantom{\mathcal{L}}}^\text{p} \tilde{\mathcal{L}}^{\Lambda_3} + {\vphantom{\mathcal{L}}}^\text{g} \tilde{\mathcal{L}}^{\Lambda_3}_\text{EH} + {\vphantom{\mathcal{L}}}^\text{g} \tilde{\mathcal{L}}^{\Lambda_3}_\text{MG}, \label{eq_action_transformed_DL_3}
\end{equation}
with
\begin{align}
	{\vphantom{\mathcal{L}}}^\text{g} \tilde{\mathcal{L}}^{\Lambda_3}_\text{EH} &= \partial_\mu C_0 \left\{ \frac{3}{2} \partial^\mu C_0 + \partial_\nu D_0 (\eta^{\mu \nu} (\partial \phi)^2 - \phi^{,\mu} \phi^{,\nu}) \right. \nonumber \\
	&\qquad + \left. D_0 (\phi_{,\nu} \Pi^{\mu \nu} - \phi^{,\mu} [\Pi]) \vphantom{\frac{3}{2}} \right\}, \\
	{\vphantom{\mathcal{L}}}^\text{g} \tilde{\mathcal{L}}^{\Lambda_3}_\text{MG} &= -A_0(X,\phi) \eta^\mn \bar{X}^{(N)}_{\mu \nu} (\pi).
\end{align}

\subsubsection{Searching for Ghosts}

The DL of the generalized disformal case is given by the purely disformal case plus a new term ${\vphantom{\mathcal{L}}}^\text{g} \tilde{\mathcal{L}}^{\Lambda_3}_\text{EH}$ from the transformed Einstein-Hilbert term and a new term ${\vphantom{\mathcal{L}}}^\text{g} \tilde{\mathcal{L}}^{\Lambda_3}_\text{MG}$ from the interaction terms.

For the first of these new terms, we use the findings of Ref. \cite{Zumalacarregui:2013pma} in which the authors argue that a general DT of the Einstein-Hilbert term does still possess a second order evolution. This implies that this term is Ostrogradsky stable.

The second of these new terms is given as a linear combination of the following terms:
\begin{align}
	A_0 \eta^{\mu \nu} X^{(1)}_{\mu \nu} &= 3A_0 [\Pi], \\
	A_0 \eta^{\mu \nu} X^{(2)}_{\mu \nu} &= A_0 ([\Pi]^2 - [\Pi^2]), \\
	A_0 \eta^{\mu \nu} X^{(3)}_{\mu \nu} &= A_0 (-[\Pi]^3 + 3[\Pi][\Pi^2] - 2[\Pi^3]).
\end{align}
These terms exhibit an interesting structure. First, by choosing $A_0(X,\phi) = X$, we obtain the Galileon Lagrangians. Second, all of these terms can be reproduced by the usual Horndeski Lagrangians with the choices
\begin{align}
	G_3(X,\phi) &= 3 A_0(X,\phi), \\
	G_{4,X} (X,\phi) &= - \frac{1}{2} A_0(X,\phi), \\
	G_{5,X} (X,\phi) &= - 3 A_0(X,\phi),
\end{align}
which are problematic, however. While the first choice works absolutely fine, the second and third one introduce new terms from the Horndeski Lagrangians, namely \\ $G_4(\phi,X) R$ and $G_5(\phi,X) G_\mn \phi^\mn$. These new terms are not included in the DL of beyond dRGT and hence further investigation is required whether the GDC is ghost-free or not.

\section{Summary and Discussion}

We investigated the possibility of a ghost-free MG that is different from the dRGT case but still has only five DoFs. For that reason, we constructed a new theory through a DT of the metric in the perturbative formulation of dRGT. Since the scalar mode $\pi$ is almost solely responsible for the phenomenology and theoretical consistency, we investigated the $\Lambda_3$-DL. However, the transformed perturbation $\tilde{h}$ of the metric around flat space-time possesses terms depending on the Planck mass $\Mp$, which give rise to divergences in any DL where the latter one goes to infinity. Therefore, we imposed that $\tilde{h}$ does not depend on the Planck mass. This was achieved by giving the free functions $C$ and $D$ of the DT an $\Mp$-dependence, which introduced new constraints. We investigated three special cases: a trivial one, a purely disformal one, where the conformal function $C$ is set to unity and a generalized disformal one, which encompasses the second one as a special case. For the PDC, we proved that the DL lives inside the class of BH Lagrangians. This proves the absence of ghosts in the DL and hints at their absence in the whole theory. For the GDC, we calculated the DL and obtained the purely disformal DL plus some extra terms. Some of these were shown to live inside the class of Horndeski Lagrangians but with problematic consequences.

In conclusion, the possibility of another way than dRGT theory to formulate a ghost-free theory for a massive spin-2 field is an exciting result that invites to future investigations. First, it would be interesting to generalize the constraints on $C$ and $D$ and thus the whole theory. This can be done by checking if the GDC lies in the class of BH, by solving the whole Eq. \eqref{eq_constraints_on_C_and_D} or by making a new ansatz as, for instance, considering the behavior of $C$ and $D$ only in the limit $\Mp \rightarrow \infty$ or allowing for an $m$-dependence. Second, apart from its Ostrogradsky stability, it is crucial to investigate the causal behavior and the possibility of non-local effects. Furthermore, when investigating the full theory for stability, one should also consider the vector modes, which have been set to zero in this analysis. Although they are usually not responsible for instabilities and there is a priori no reason to expect otherwise, they should be included for the sake of completeness. Finally, the identification of the disformal field with the helicity-0 mode $\phi = \pi$ breaks the $U(1)$-symmetry of the helicity-0 mode, which has been used in dRGT in order to preserve the DoFs. Since our findings imply that the DoFs in beyond dRGT are probably conserved, we assume that there is another, hidden mechanism in it which plays the same role as the $U(1)$-symmetry and thus preserves the DoFs.
\par
Shortly after this manuscript has been submitted, Ref. \cite{Golovnev:2018icm} has appeared, in which the author states that a subclass of beyond dRGT can be interpreted as a mimetic extension of dRGT. Since a mimetic theory does not introduce a new propagating DoF, it supports the claim that beyond dRGT is actually ghost-free.

\phantom{\rule{\textwidth}{1pt}}

\section*{Acknowledgments}
We would like to thank Miguel Zumalac\'{a}rregui for his constantly invaluable advice. Furthermore, our thanks go to Angnis Schmidt-May for helpful comments and interesting discussions. M.W. thanks the DFG for support through the Research Training Group ``Particle Physics beyond the Standard Model'' (GRK 1940). Furthermore, L.A. acknowledges support from the DFG through the TRR33 project ``The Dark Universe''. Finally, N.K. thanks the School of Physics at IPM where he is a part-time researcher and the ITP at Heidelberg University for supporting his visit when this project was initiated.

\bibliographystyle{elsarticle-num}
\biboptions{sort&compress}
\bibliography{references}

\end{document}